\documentclass[seceq]{ptptex}

\usepackage{graphicx}

\usepackage{dcolumn}
\usepackage{bm}

\newcommand{\TP}{\Theta^{+}}
\makeatletter
\def\simleq{\mathrel{\mathpalette\gl@align<}}
\def\simgeq{\mathrel{\mathpalette\gl@align>}}
\def\gl@align#1#2{\lower.6ex\vbox{\baselineskip\z@skip\lineskip\z@
\ialign{$\m@th#1\hfill##\hfil$\crcr#2\crcr\sim\crcr}}}
\makeatother

\newcommand{\bra}{\langle}
\newcommand{\ket}{\rangle}
\newcommand{\braket}[1]{\bra #1 \ket}

\newcommand{\qq}{\braket{\bar{q}q}}

\newcommand{\sbs}{\braket{\bar{s}s}}
\newcommand{\qGq}{\braket{\bar{q}g_s \sigma\cdot G q}}

\newcommand{\sGs}{\braket{\bar{s}g_s \sigma\cdot G s}}
\newcommand{\GG}{\braket{\frac{\alpha_s}{\pi} G^2}}

\def\Eq.#1{Eq.~(\ref{#1})}

\newcommand{\MB}{M}

\newcommand{\half}{{1\over 2}}

\newcommand{\sth}{\sqrt{s_{\rm th}}}

\newcommand{\tenbar}{\overline{10}}

\notypesetlogo                       

\title{
Theoretical Overview of the Pentaquark Baryons%
}

\author{
Makoto \textsc{Oka}%
}

\inst{
Department of Physics, H27, Tokyo Institute of Technology\\
Tokyo 152-8551, Japan
}

\recdate{May 29, 2004}

\abst{
A review is given of the theoretical ideas concerning the mysterious pentaquark baryons proposed during the first year after its discovery.  We focus on the difficulties involved with the constituent quark models and the discrepancy between the QCD predictions and most of the phenomenological models.
}

\begin{document}

\maketitle

\section{Introduction}

The year 2003 will be remembered as a renaissance of hadron spectroscopy.
Indeed, there were many reports of the discovery of new narrow resonances.
The list includes the celebrated pentaquark baryons,\cite{Nakano} 
the new charmonium state $X(3872)$,\cite{Belle} new $D_s$ mesons,\cite{BaBar} 
and even a narrow $K$ bound 
nucleus.\cite{Iwasaki}  
Among these the discovery of the first pentaquark  baryon, $\TP$, was the most striking event.

The exotic baryon $\TP$ was discovered in a seminal experiment performed by the LEPS 
group at the SPring-8 facility in Harima, Japan\cite{Nakano}. 
This particle was identified in 
the $K^+n$ invariant mass spectrum 
in the reaction $\gamma n \to  K^- + \Theta^+ \to K^- +K^+ + n$,
which was induced by a SPring-8 tagged photon beam of energy up to 2.4 GeV.
That search was motivated by a theoretical paper by Diakonov, Petrov and Polyakov
 (DPP)\cite{DPP}, in which they predicted a narrow $S=+1$ baryon.

The reported mass of $\TP$ is $1540\pm 10$ MeV, and the upper limit of
its width is 25 MeV.
As this baryon has strangeness $+1$, it should contain at least one $\bar s$, and
thus the minimal number of quarks for $\TP$ is five. From the charge
and the strangeness, $u^2d^2\bar s$ is a possibility as the content of $\TP$. It is therefore 
commonly called the `pentaquark'.

The discovery of $\TP$ inspired many other experiments, and the existence of $\TP$ 
was soon confirmed
by various groups, including ITEP (DIANA),\cite{ITEP} JLAB (CLAS)\cite{CLAS} 
and ELSA (SAPHIR).\cite{SAPHIR}
Its discovery was followed by the discovery of 
yet another exotic baryon, $\Xi^{--}$, found by the NA49 group at CERN\cite{NA49}.
The particle $\Xi^{--}(s^2d^2\bar u)$ is another manifestly exotic baryon, 
whose decay into $\Xi^- \pi^-$ has been
observed at the mass
$M = 1.862$ GeV with a width $\Gamma < 18$ MeV.

Since the discovery of $\TP$ was reported, numerous papers attempting to understand its properties and structure have been uploaded to the preprint server. 
In this article, we pick out some of the prominent achievements among them, 
attempt to explain their content, and also point out some problems.
We concentrate on $\TP$, first summarizing the experimental facts (\S 2) and
then giving an overview of several important theoretical ideas
(\S\S 3 and 4).
In particular, we discuss the predictions of the quark models with regard to the pentaquark states.

As QCD is the fundamental theory of hadrons, our goal is to understand
$\TP$ from the viewpoint of QCD.  Several attempts have been made using the QCD sum rule and lattice QCD simulations. 
In \S 5, we give an overview of the results and their interpretation.
We give a conclusion in \S 6.

\section{Facts}

A series of experiments have confirmed that the mass of $\TP$ is
$M_{\TP}\sim 1540$ MeV, while the upper limit of the width is 9 MeV, 
although the accumulated data exhibits some dispersion of the central value of the
mass.\cite{EXP_summary}
There is no consensus regarding the width, although
the HERMES group\cite{HERMES} claims a non-upper-limit width of order 10 MeV.
 (Their mass value is somewhat smaller, 1528 MeV.)
Because it decays into $nK^+$ by the strong interaction, the conservation rules guarantee
that it has a strangeness $S=+1$,  baryon number $B=+1$, and charge $Q=+1$.
Thus, the hypercharge is $Y\equiv B+S =2$, and the third component of the isospin is 
$I_3=0$. 
No corresponding $pK^+$ ($I_3= +1$) state is observed at the same mass.
This leads to the conclusion that the isospin of $\TP$ is $I =0$ (see Table I). 
It is obvious that this state cannot be composed of only three quarks.
It also seems important that no $S = +1$ baryon state has been observed
below the $NK$ threshold, and thus this state seems to be the ground state.

As the flavor $SU(3)$ is an approximate symmetry of QCD, it is likely that $\TP$ 
belongs to an $SU(3)$ multiplet.
The simplest $SU(3)$ irreducible representation (IR) for $\TP$ is $\tenbar$, which is, in fact, unique if we restrict ourselves to the representations produced by 5 quarks (Fig.~1).
This is the representation obtained from the Skyrmion model, and also it
is that assumed in most of the quark model descriptions.
We therefore assume here that $\TP$ is a member of $\tenbar$.

\begin{table}[htdp]
\begin{center}
\begin{tabular}{cc|cccc|l}
$I$&$I_3$&$B$&$S$&$Y$&$Q$&$J^{\pi}$\\
\hline
0&0&1&1&2&1&$1/2^+,1/2^-, \ldots$\\
\end{tabular}
\end{center}
\caption{The quantum numbers of $\TP$. The spin and parity have not been determined.}
\label{tab:quantum_numbers}
\end{table}

\begin{figure}[htbp]
\begin{center}
\includegraphics[width=12cm]{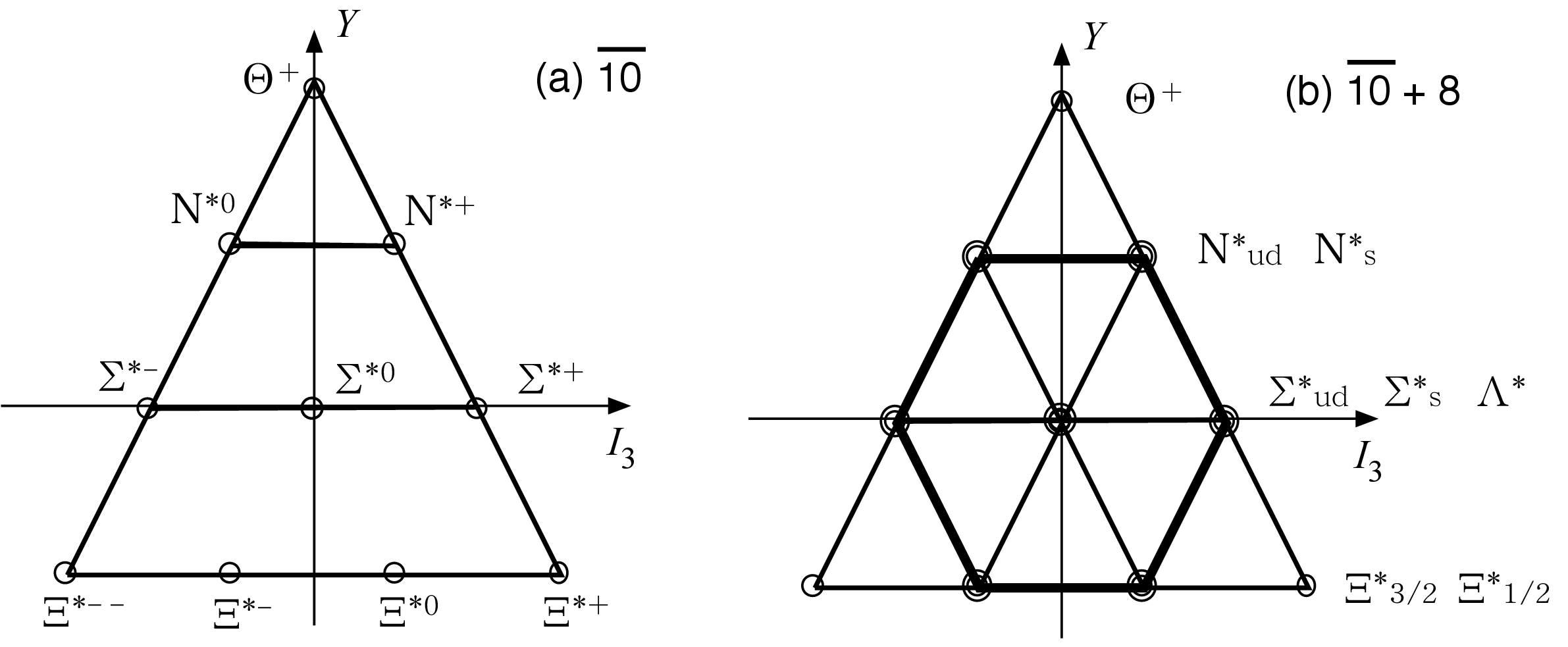}
\caption{(a) The $\tenbar$ pentaquarks and (b) the mixing of the $\tenbar$ plus 8 representations for the pentaquarks.}
\label{10}
\end{center}
\end{figure}

Where are the other members associated with $\tenbar$?
An important member is the penta-quark $N^*$ ($I =1/2$, $Y =1$), 
which is the state with the flavor content:
\begin{equation}
N_{\overline{10}}^*= [(2/3) \bar s s + (1/3) (\bar u u+ \bar d d)] 
\times (uud \hbox{ or } ddu) .
\label{eq:N10}
\end{equation}
Because this is a nucleon resonance, it is quite plausible that 
the present PDG table\cite{PDG} already includes this state.
DPP\cite{DPP} boldly identified a $1/2^+$ nucleon resonance, $N$(1710), 
as one such possibility. 
It is, however, noted that the $SU(3)$ breaking can cause this state to be mixed with other representations.
Jaffe and Wilczek (JW)\cite{JW} proposed the ideal mixing of a pentaquark octet with 
$\tenbar$, that is, they predicted the two nucleon states
\begin{eqnarray}
N_{ud}^* &=& (\bar u u+ \bar d d) \times (uud \hbox{ or } ddu), \nonumber\\
N_s^* &=& \bar s s \times (uud \hbox{ or } ddu) .
\end{eqnarray}
JW further identified these two states as the two lowest $1/2^+$ resonances, 
$N$(1440) and $N$(1710).
It may, however, be possible that they correspond to, for example,
the $1/2^-$ states $N$(1535) and $N$(1650). 
If so,  it is possible that $\TP$ has negative parity.

\begin{table}[htdp]
\begin{center}
\begin{tabular}{l|r|c|c}
\hline\hline
Diquarks&$J^{\pi}$ & color& flavor\\
\hline\hline
$\bar S_a^5 \equiv \epsilon_{abc}(u_b^T C \gamma^5 d_c) $& $0^+$ & $\bar 3$& $\bar 3$\\
$\bar S_a \equiv \epsilon_{abc}(u_b^T C  d_c) $& $0^-$ & $\bar 3$& $\bar 3$\\
\hline
$\bar U_a^5 \equiv \epsilon_{abc}(d_b^T C \gamma^5 s_c) $& $0^+$ & $\bar 3$& $\bar 3$\\
$\bar U_a \equiv \epsilon_{abc}(d_b^T C  s_c) $& $0^-$ & $\bar 3$& $\bar 3$\\
\hline
$\bar D_a^5 \equiv \epsilon_{abc}(s_b^T C \gamma^5 u_c) $& $0^+$ & $\bar 3$& $\bar 3$\\
$\bar D_a \equiv \epsilon_{abc}(s_b^T C  u_c) $& $0^-$ & $\bar 3$& $\bar 3$\\
\hline\hline
Pentaquarks&$J^{\pi}$ &  flavor& \# of $s, \bar s$\\
\hline\hline
$\TP = \epsilon_{abc} \bar S_a \bar S^5_b  C\bar s^T_c$& $1/2^+$ &  $\tenbar$& 1\\
$\TP_{NK} = \epsilon_{abc}([u_a^T C\gamma^5 d_b] u_c (\bar s_d\gamma^5 d_d) +
(u\leftrightarrow d) )$& $1/2^-$ & $\tenbar$& 1\\
\hline
$N_{ud}^* = \epsilon_{abc} \bar S_a \bar S^5_b  C\bar u^T_c$& $1/2^+$ &  $\tenbar+8$& 0\\
$N_s^* = \epsilon_{abc} \frac{1}{\sqrt{2}}
(\bar U_a \bar S^5_b  +\bar S_a \bar U^5_b  ) C\bar s^T_c$& $1/2^+$ &  $\tenbar+8$& 2\\
$N_{\tenbar}^* = (2/3) N_s^* + (1/3) N_{ud}^*$ & $1/2^+$ &  $\tenbar$& 4/3\\
\hline
$\Sigma_{ud}^* = \epsilon_{abc} \frac{1}{\sqrt{2}}
(\bar U_a \bar S^5_b  +\bar S_a \bar U^5_b)  C\bar u^T_c$& $1/2^+$ &  $\tenbar+8$& 1\\
$\Sigma_s^* = \epsilon_{abc} 
\bar U_a \bar U^5_b C\bar s^T_c$& $1/2^+$ & $\tenbar+8$& 3\\
$\Sigma_{\tenbar}^* = (2/3) \Sigma_s^* + (1/3) \Sigma_{ud}^*$ & $1/2^+$ 
&  $\tenbar$& 5/3\\
\hline
$\Xi_{3/2} = \epsilon_{abc} \bar U_a \bar U^5_b  C\bar u^T_c$& $1/2^+$  & $\tenbar$& 2\\
\hline\hline
\end{tabular}
\caption{The local QCD operators for diquarks and pentaquarks. The color is denoted by $a$, $b$ and $c$.  $C=i\gamma^2\gamma^0$ is the charge conjugation matrix.  $\TP_{NK}$ is the product of $N$ and $K$ with isospin $I=0$.}
\end{center}
\label{tab:pentaquark}
\end{table}

A similar ideal mixing is plausible for $\Sigma$ ($S=-1$) states, while $\Xi$ ($S=-2$) appear
in two multiplets, $\Xi^*$ ($I=3/2$) in $\tenbar$ and $\Xi^*$ ($I=1/2$) in 8.
They do not mix with each other because of the isospin symmetry.
The effective number of the strange quarks, $n_s$, is the key to understanding the $\tenbar$ structure and its mixing with 8.
In Table II, we list the flavor components of the pentaquark states in the form of the local QCD operators.
We use ``diquark'' notation, in which two quarks are combined into the color $\bar 3$
(antisymmetric), the flavor $\bar 3$  (antisymmetric) and the spin 0 (antisymmetric) states.
It is seen from Eq.~(\ref{eq:N10}) that the average number of strange quarks is 1 for $\TP$, and
it increases to $n_s=4/3$ for $N^*$,
$n_s=5/3$ for $\Sigma^*$, and $n_s=2$ for $\Xi^*$ in the $\tenbar$ representation.

It is easy to see that the maximal breaking of $SU(3)$ leads to the ideal mixing, where 
$n_s(N_{ud}^*) = 0$, while  $n_s(N_{s}^*) = 2$. 
Similarly, $n_s(\Sigma_{ud}^*) = 1$ and $n_s(\Sigma_{s}^*) = 3$.
Thus, after the ideal mixing, the spectrum should appear as $N_{ud}$, ($\TP$, $\Sigma_{ud}^*$),
($N_s^*$, $\Xi_{3/2}^*$, $\Xi_{1/2}^*$) $\Sigma_s^*$, where the multiplets in the parentheses have the same value of $n_s$ and are degenerate to leading order in $m_s$.
JW predicted the masses of the $\Xi_{3/2}^*$ and $\Xi_{1/2}^*$ states to 
be around 1750 MeV, while the original result of DPP predicts 
a much larger mass for $\Xi_{3/2}^*$.

\section{Theory I: Soliton model}

\subsection{Skyrmion}

The search for the $S=+1$ pentaquark at SPring-8 was motivated by a theoretical prediction of
DPP.\cite{DPP}
That study is based on the Skyrme soliton (Skyrmion) model of the baryon,\cite{Skyrmion}
 which utilizes 
a topological soliton solution in the $SU(3)\times SU(3)$ chiral symmetric nonlinear 
sigma model. 
Their model has specific $SU(3)$ breaking terms, indicated by the chiral quark model, but
the strengths of the individual terms of the effective Lagrangian are determined phenomenologically.

In the Skyrme model, the classical soliton solution is invariant under neither $SU(3)$ nor
spatial rotations.  The canonical form is the $SU(2)$ hedgehog imbedded in 
$U\in SU(3)$:
\begin{equation}
U(\vec r) =\exp (i\vec\lambda\cdot\hat{r} F(r)),
\end{equation}
where
$$ \vec\lambda \equiv (\lambda_1, \lambda_2, \lambda_3) = 
\begin{pmatrix}
\vec\tau & 0 \cr 0 & 0 \cr
\end{pmatrix} . $$
The profile function $F(r)$ represents the radial shape of the soliton and satisfies the boundary conditions $F(r=0) =\pi$ and $F(r\to\infty)=0$, so that the topological winding number is equal to 1, which is identified with the baryon number.
To recover the symmetries and thus to specify the
spin and flavor quantum numbers, we need to ``rotate'' the Skyrmion in $SU(3)$ flavor space
and project the appropriate quantum numbers.
This method of quantization is called the `rigid rotor quantization'.
In the $SU(3)$ Skyrme model with $N_c=3$, the lowest-lying 
states are the flavor octet and decuplet states, which correspond to the
observed ground state baryons\cite{SU3Skyrmion} and are
also consistent with the quark model prediction 
with three quarks in the baryon. 
The next possible representation, $\tenbar$, 
is not composed of three quarks and therefore is exotic.  
Its spin and parity are restricted to 
$1/2^+$ in this model.

DPP assumed that the 
known nucleon resonance $N^*$(1710) is the $Y=1$ member of the 
$\tenbar$ multiplet and
derived the (Gell-Mann--Okubo) mass formula
\begin{equation}
M = (1890 - 180 \,Y ) \hbox{ MeV} ,
\end{equation}
using the parameters of the Hamiltonian determined so as to reproduce the octet and
decuplet baryons.
With this formula, the mass of the $Y=2$ member ($B=1$ and $S=+1$) is predicted to be
$M =1530 \hbox{ MeV}$.
This member was first called $Z^*$ but it is now called $\TP$.
They also calculated the width of $\TP$ in the Skyrme model, assuming the simplest 
meson baryon coupling operator. They found that its width may be as small as 15 MeV.%
\footnote{Subsequently, it was pointed out that there is a numerical error in their calculation, and the correct value is 30 MeV.\cite{Jaffe}}

It is important to note that the Skyrme model predicts only positive parity states, unless
non-collective meson fluctuations are introduced around the Skyrmion.  There is another constraint, namely that the spin is also specified by the $SU(3)$ representation.
This constraint is imposed by the Wess-Zumino-Witten term; that is,
it results from the stipulation that 
the allowed representation contains a member with ($I=J$, $Y=1$).
Thus, the spin should be equal to the isospin of the $Y=1$ member of 
the specified representation.\cite{SU3Skyrmion}
Hence, $J^{\pi}= 1/2^+$ is the only allowed spin for the $\tenbar$ flavor state.

There are additional studies based on the Skyrmion model.  Their main
achievement is to take account of the mixings of other representation due to the $SU(3)$ breaking
and to fit the parameter values accordingly.  Yabu and Ando\cite{YA} carried out 
a full study of the representation mixing in the octet and decuplet baryons some time ago, and
the basic idea is given in their paper.
In the case of $\tenbar$, mixings of 8 and 27 have been introduced, but the qualitative features do not change as a result of the mixings\cite{Pra}.
Diakonov and Petrov\cite{DP2} have reexamined their study after the discovery of $\Xi^*$, finding that the choices of the parameter values can be changed so that a smaller mass 
of $\Xi_{3/2}^*$ is obtained.

We would like to point out here that the Skyrme model in general predicts 
many (sharp) resonances, which may not necessarily correspond to real baryons,
because the model is assumed to represent QCD at large $N_c$.  
There is a consensus that there is a significant
$1/N_c$ correction, especially for the excited states.
Furthermore, several authors have pointed out that the rigid rotor quantization for the
$\tenbar$ representation is not consistent with the large $N_c$ 
expansion\cite{Cohen,IKOR}.

\subsection{$K^+$ Skyrmion bound state approach}

Itzhaki et al.\cite{IKOR} studied an alternative quantization of the $SU(3)$ Skyrmion,
the kaon-Skyrmion bound state approach, and they 
compared their method with the rigid rotor quantization.
This alternative was originally proposed by Callan and Klebanov\cite{CK} in the
context of describing the strange baryon octet and decuplet.
They pointed out that the WZW term is repulsive (attractive) for $K^+$ ($K^-$),
and the $K^+$ bound state does not exist in their original work.
Itzhaki et al.\cite{IKOR} reexamined this case and found that
if the kaon mass is as large as 1 GeV, then a bound state may appear.
In that case, the most probable quantum number is $1/2^+$.

Hosaka\cite{Hosaka} pointed out that the chiral bag model exhibits similar 
qualitative features and predicts a positive parity ground state for $\TP$.
The chiral bag model has a solution of hedgehog form and, in a sense, interpolates 
between the two extreme pictures, the quark model (MIT bag model) and the Skyrmion model.
Hosaka considered a configuration of 4 $ud$ quarks and 1 $\bar s$
as a function of the strength of the pion field at the surface of the bag.
The first three light quarks occupy the lowest $K=0$ state.
($K$ is the grandspin defined by $\vec K\equiv \vec I + \vec J$.) 
The $\bar s$ state is almost independent of the pion field.
The fourth light quark is found to occupy the negative parity $K^{\pi} =1^{-}$ state
if the surrounding pion field is sufficiently strong.
This happens because the energy of the $K^{\pi} =1^{-}$ state goes below the energy of
the $K^{\pi} =1^{+}$ state for a stronger pion field.
In this case, the total spin and isospin are
$$ uudd (I =0, J^{\pi} =1^-)  + \bar s (J^{\pi} =1/2^{-}) \to J^{\pi} = 
1/2^+ \hbox{  and } 3/2^+ .$$
This mechanism is qualitatively consistent with the kaon-Skyrmion bound state.

The kaon bound state, or the chiral bag model, may be more appropriate than the rigid-rotor quantized Skyrmion.  But these two approaches cannot be extended to all the members of $\tenbar$. Thus we need new ideas to describe the $\Xi^*$ pentaquark.

\section{Theory II: Quark Models}

As seen above, the predictions obtained from the Skyrmion model apply only to states with limited quantum numbers. In contrast, QCD with quarks and gluons should have a richer spectrum with various spin and flavor quantum numbers. We therefore consider the general quark model in this section and find what it can yield for the pentaquark states.

\subsection{Symmetry in quark models}

First we consider the symmetry restrictions on the tetra-quark states
composed of $u^2 d^2$ quarks.
We consider only the $I=0$ and $C=3$ states, so that they are a part of the $\TP$ baryon.
With four quarks, this state possesses the following symmetry:\footnote{%
We here consider the case for $\TP$, but the same symmetry classification applies to other pentaquark states. For the other states, we replace $I=0$ by the flavor representation, $F = \bar 6$.}
$$ I=0\quad 
\begin{picture}(2,1) \multiput(0,1)(1,0){2}{\framebox(1,1){}}
\multiput(0,0)(1,0){2}{\framebox(1,1){}}
\end{picture} 
\quad \otimes\quad C=3\quad 
\begin{picture}(2,2) \multiput(0,1)(0,-1){3}{\framebox(1,1){}}
\put(1,1){\framebox(1,1){}}
\end{picture} 
\quad =\quad 
\begin{picture}(2,2) \multiput(0,1)(0,-1){3}{\framebox(1,1){}}
\put(1,1){\framebox(1,1){}}
\end{picture} 
\quad\oplus\quad
\begin{picture}(3,2) \multiput(0,1)(1,0){3}{\framebox(1,1){}}
\put(0,0){\framebox(1,1){}}
\end{picture} 
$$

\vskip 12pt
The remaining degrees of freedom corresponds to the spin and orbital motion.
For simplicity, we consider the classification in the nonrelativistic
limit.
If we assume that the ground state is the orbitally symmetric state with $L=0$,
then the only allowed spin is $S=1$ with [31] symmetry.

However, the conventional quark model yields the conclusion that the spin dependent interactions,
especially the hyperfine interaction (HF), are important. The HF favors lower spin states, 
and therefore the $S=1$ tetraquark state is less favored.
On the other hand, the $S=0$ state is allowed only with the mixed orbital symmetry, [31],
having $L=1$.
Therefore the combined total angular momentum of the tetraquark is $j=1$.
Such a state is favored by the HF interaction, but the cost is its greater kinetic energy.

Thus, the possible low energy configurations for the $\TP$ baryon are
\begin{eqnarray*}
&\hbox{(A) }&u^2d^2(L=0, S=1, j^{\pi}=1^+) + \bar s (1/2^-) : J^{\pi}  
= 1/2^- \hbox{ or } 3/2^- , \\
&\hbox{(B) }&u^2d^2(L=1, S=0, j^{\pi}=1^-) + \bar s (1/2^-) : J^{\pi}  
= 1/2^+ \hbox{ or } 3/2^+ .
\end{eqnarray*}
In either case, there should exist a spin partner, $J=3/2$
state, associated with the ``ground'' $J=1/2$ state.%
\footnote{Such a state with the same flavor symmetry does not necessarily
exist in the Skyrmion picture.}
However, the origin of the $J=1/2-3/2$ splitting differs.
For case (A), the spin-spin interaction,
$\propto \vec S(u^2d^2)\cdot \vec S(\bar s)$,
 is responsible for splitting them,
while for case (B), the (multibody) spin-orbit force, 
$\propto \vec L(u^2d^2)\cdot\vec S(\bar s)$,
is the interaction which causes the splitting.
In the ordinary hadron spectrum, the spin-spin splitting is much larger than
the spin-orbit splitting. Furthermore the ``three-body'' LS force in
case (B) is expected to be smaller than the two-body LS.
According to Dudek and Close\cite{Close}, the splitting in case (B) is 
of the order of a few tens
of MeV, while the spin-spin splitting in case (A) is $\sim 200$ MeV.
In any case, the observation of the $J=3/2$ partner of $\TP$ would be critical
in oder to 
distinguish among the models of the structure of the pentaquark baryons.

We consider the above two cases further in order to determine which of the two is
preferable in the constituent quark model approaches.
The main question is which of the following two competing energies is larger:
\begin{itemize}
\item Orbital excitation energy: 
$\Delta E(\Delta L=1) = E(L=1) - E(L=0)$,
\item Hyperfine (HF) interaction:
$\Delta E(\Delta S=1) = E(S=1) - E(S=0)$.
\end{itemize}
If $\Delta E(\Delta L=1) > \Delta E(\Delta S=1)$, case (A) with $J^{\pi}=1/2^-$ may give
the ground state, while if $\Delta E(\Delta L=1) < \Delta E(\Delta S=1)$,
case (B) with the $1/2^+$ ground state may be realized.

\subsection{Constituent quark model}

A simple constituent quark model is employed here to estimate the above
two quantities using the knowledge we have regarding the ordinary hadron spectrum.
The constituent quark is known to be a very useful concept in describing the meson
and baryon spectra and their properties.  It has a mass of order 300 MeV,
which is believed to be generated by the chiral symmetry breaking in QCD.
It can be derived by solving the Dyson-Schwinger equation for the quark 
propagator\cite{BS}. The interaction kernel has strong enhancement in the low momentum
region, and it induces dynamical chiral symmetry breaking and generates a momentum
dependent effective mass.
Thus the quark acquires the self-energy and is renormalized. 
It is important to note that the conserved currents, such as
the isospin, $I$, the hypercharge, $Y$, and color charge, $C$,
are not renormalized.

Therefore, the constituent quark has the same quantum numbers as the QCD quark
but its mass is given roughly by $M_B/3$.  
We choose, as typical values of the quark mass,
$m_{\rm u, d} \sim 360$ MeV and $m_s \sim 540$ MeV.
Then the sums of the quark masses for the baryons become
\begin{eqnarray}
\sum m_q &=& 
\begin{cases}
1080 \hbox{ MeV for $N$ and $\Delta$,}\cr
1260 \hbox{ MeV for $\Lambda$, $\Sigma$ and $\Sigma^*$,}\cr
1980 \hbox{ MeV for $u^2d^2\bar s$.}\cr
\end{cases}
\end{eqnarray}
Assuming that the residual interactions are weak,%
\footnote{A confinement force is, of course, required to combine quarks into bound states.
This effect is assumed to be within the constituent quark mass; i.e., $m_q$ is regarded as
the single particle energy of the quarks in the ground state.} 
these values provide estimates of the average masses of the indicated baryons.

\subsection{Orbital excitation}

We first estimate the orbital excitation energy, that is, the energy 
necessary for 
the excitation of a constituent quark from the lowest single particle 
energy level to the first excited level with $L=1$.
This energy can be estimated from the excitation energy of the negative-parity
baryon resonance, e.g., 
$N^*(1535) - N(940)$, $\Delta E \sim 600$ MeV.
This is, however, an overestimate, because it contains the difference of the hyperfine interactions
in the two baryons.

A more reliable estimate may be obtained by calculating the difference of the kinetic energies in
the quark-diquark system.  For quarks of constituent mass $\sim 360 $ MeV, the reduced mass becomes
\begin{equation}
\mu = 360\times\frac{3}{4}= 270 \hbox{ MeV.}
\end{equation}
Assuming that the distance between the quark and the diquark is about 0.6 fm, which is a typical size of the quark wave
function in the hadron, we obtain
$\Delta E(\Delta L=1) \sim 400$ MeV.

It is interesting that the JW diquark model may require even a larger
kinetic energy excess, because, if we assume that the diquark mass is about 420 MeV,
then the reduced mass for the diquarks is 210 MeV.

\subsection{Hyperfine interaction}

The importance of the hyperfine interaction in the hadron spectrum was first pointed out 
by DeRujula, Georgi and Glashow\cite{DGG} in the context of one-gluon exchange.
It is the spin-spin interaction given by the magnetic component of the one-gluon 
exchange (Fig.\ 2(a)) and is written in nonrelativistic form as
\begin{equation}
\frac{\alpha_s}{m_i m_j} (\lambda_i \cdot\lambda_j)  (\vec\sigma_i \cdot\vec\sigma_j)  
\delta (\vec r_{ij}) .
\end{equation}
This is known as the color magnetic (CM) interaction.
Its main contribution in the baryon spectrum is to split the octet and decuplet baryons, such as $N$ and $\Delta$.

We simplify the form of the CM as 
\begin{equation}
	\Sigma_{\rm CM}  =  \sum_{i<j} \Delta_{\rm CM} \,
	\xi_{ij}\,(\lambda_i\cdot\lambda_j)
	(\sigma_i\cdot\sigma_j)  
\end{equation}
and fix the strength $\Delta_{\rm CM}$ using the $N-\Delta$ mass difference, which
leads to
\begin{equation}
\Delta_{\rm CM} = 18.75 \hbox{ MeV.}
\end{equation}
Here $\xi$ denotes a factor due to the $SU(3)$ symmetry breaking, defined by
\begin{equation}
\xi_{ij}= \begin{cases}
1& \hbox{for $(ij) = (uu)$, $(ud)$ or $(dd)$,}\cr
m_u/m_s & \hbox{for $(ij)=(us)$ or $(ds)$,}\cr
(m_u/m_s)^2 & \hbox{for $(ij)=(ss)$.}\cr
\end{cases}
\end{equation}
This is introduced so that the mass splitting ($\sim 80$ MeV) of $\Lambda$ and $\Sigma$
is reproduced by the CM interaction if we take $m_s/m_u \sim 5/3$.

\begin{figure}[htbp]
\begin{center}
\includegraphics[width=13cm]{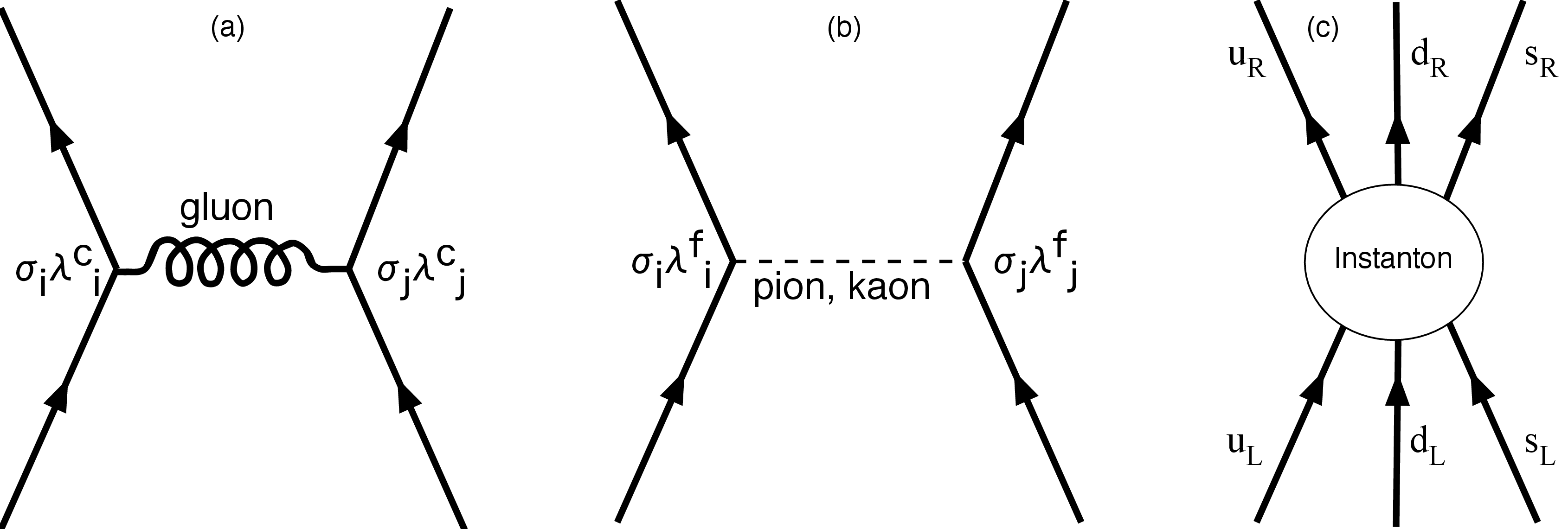}
\caption{(a) CM from the one gluon exchange, (b) FD from the exchange of the pseudoscalar mesons and (c) III due to the quark-instanton coupling.}
\label{OgEetc}
\end{center}
\end{figure}

Other possible origins of the HF interaction have been discussed in the literature.
One of the approaches predicts pseudoscalar meson exchange interactions between quarks inside the baryon (Fig.\ 2(b)).	We call this interaction the ``flavor dependent" (FD) HF interaction\cite{Glozman}.
This force depends on the flavor and spin of the quarks and is given in simplified form as
\begin{equation}
\Sigma_{\rm FD} = - \Delta_{\rm FD} \sum_{i<j} \xi_{ij}
(\lambda^f_i \cdot\lambda^f_j) (\vec\sigma_i\cdot\vec\sigma_j)
\end{equation}
The strength can be determined again using the $N-\Delta$ mass difference:
$\Delta_{\rm FD} \sim 30$ MeV.

\def\Viii{V_{\rm I\,I\,I}}
\def\vecsig{\vec{\sigma}}
\def\vecr{{\vec r}}
The third possible origin of the HF interaction is the instanton induced interaction (III) (Fig.\ 2(c)).  
It was shown by Shuryak and Rosner\cite{ShR} and Takeuchi and Oka\cite{TO} that the III gives the correct HF interaction for the ground-state baryon spectrum. Takeuchi\cite{Takeuchi} further showed that the spin-orbit interaction in the P-wave baryons is also described well by the III.
The III is also known as 
the Kobayashi-Maskawa-'t Hooft (KMT) interaction, and it results from the coupling of light quarks with the instanton. It is given as follows:
\begin{eqnarray}
	\Viii^{(3)}&=& V^{(3)} \sum_{(ijk)} {\cal A}^{f} \left[
	1-\frac{1}{7}\, \left( \vecsig_{i}\cdot\vecsig_{j} +
	\vecsig_{j}\cdot\vecsig_{k} +\vecsig_{k}\cdot\vecsig_{i} \right)\right]
	\delta(\vecr_{ij})\delta(\vecr_{jk}) , \nonumber\\
	\Viii^{(2)}&=& V^{(2)} \sum_{i<j} {\cal A}^{f} \left[
	1-\frac{1}{5}( \vecsig_{i}\cdot\vecsig_{j}	)  
 	\right]
	\delta(\vecr_{ij}) .
	\label{eq:Viii}
\end{eqnarray}
It contains the flavor antisymmetric ($u-d-s$) 3-body repulsion and also the flavor antisymmetric 2-body attraction, which is induced by contracting the 3-body force	by contracting a pair consisting of a quark and an antiquark.

The two-body part of the III can be simplified into the HF form,
\begin{equation}
\Sigma_{\rm III} = \Delta_{\rm III} \sum_{i<j} {\cal A}^f_{ij}\xi_{ij}
\left[1-\frac{1}{5} (\vec\sigma_i\cdot\vec\sigma_j)\right],
\end{equation}
and its strength is given by $\Delta_{\rm III} \sim 125$ MeV.

Although the III is not uniquely identified in the baryon spectrum, its importance is known in the meson spectrum and the dibaryon spectrum.
't Hooft\cite{tHooft} pointed out that the III breaks $U_A(1)$ symmetry and thus gives the $\eta$ and $\eta'$ mass difference. Indeed, the above form of the III was used by Hatsuda and Kunihiro\cite{HK} in an NJL type model to reproduce the pseudoscalar meson spectrum. 
The importance of the three-body III in the $H$ dibaryon prediction was pointed out by Takeuchi and Oka\cite{TOH}; that is, the three-body repulsion pushes the $H$ mass above the 2$\Lambda$ threshold.  

Which of the contributions of the HF is most important?  
It is quite likely that the actual HF interaction is a combination of $\Sigma_{\rm CM}$, 
$\Sigma_{\rm FD}$ and $\Sigma_{\rm III}$.
But, we would like to stress that the axial $U(1)$ symmetry breaking caused by the III 
manifests itself in the pseudoscalar mesons, $\eta$ and $\eta'$.  Thus at least a part of the HF interaction must come from the III.  This fact is also favorable, because the CM interaction can be reduced so that an abnormally large $\alpha_s$ is not necessary to explain the baryon spectrum.

\subsection{$\TP$ mass}
Now, using the above simplified forms of the HF interactions, we estimate the baryon masses and apply the same calculation to the pentaquarks.
The masses of the nucleon and the $\Delta$ resonance are given in the CM model by
\begin{eqnarray}
		M_N &=& 3 m_q + \langle \Sigma_{\rm CM}\rangle_N  = 360\times 3 - 150 = 930 
		\hbox{ MeV,} \\
		M_{\Delta} &=& 3 m_q + \langle \Sigma_{\rm CM}\rangle_{\Delta}  
		= 360\times 3 + 150 = 1230 	\hbox{ MeV.} 
\end{eqnarray}
Similarly, the $SU(3)$ breaking effects are properly taken into account in that model,
the $\Lambda$ mass is given by
\begin{eqnarray}
		M_{\Lambda} &=& 3 m_q + \Delta m + \langle \Sigma_{\rm CM}\rangle_{\Lambda}  
		= 360\times 3 +180 - 150 = 1110 	\hbox{ MeV,} 
\end{eqnarray}
where the mass difference between the $s$ quark and the $u$ and $d$ quarks is given by
\begin{equation}
\Delta m = m_s - m_q \sim 540 - 360 = 180 \hbox{ MeV.}
\end{equation}
We note again that all the spin dependent interactions are assumed to be included in the ``single particle'' energy, i.e., the constituent quark mass.

It is found that the CM is not sufficient for the $H$ dibaryon, a system of six quarks,
represented by $u^2d^2s^2$, with strangeness $-2$.
The CM predicts a deep bound state for $H$, whose mass is estimated as
\begin{equation}
		M_H = 6 m_q + 2\Delta m +  \langle \Sigma_{\rm CM}\rangle_{H} 
		= 360\times 6 + 2\times 180 -  450 = 2070 \hbox{ MeV,}
\end{equation}
while the $\Lambda\Lambda$ threshold is 2231 MeV.
Since the prediction of the bound $H$ by Jaffe\cite{Jaffe-H} in 1977, 
experimental searches were performed for twenty years with a negative result.
This is the case in which the 3-body III gives approximately a 160 MeV repulsion.\cite{TOH}
Although the other baryons are reproduced fairy well by any of the three models, 
the fact that the $H$ dibaryon has not been observed gives reason to believe that the instanton is necessary.

Now we evaluate the mass of the $\TP$ using the simple model described above.
For the negative parity $\TP$ ($1/2^-$), we obtain
\begin{equation}
	   M_{\TP} = 5 m_q + \Delta m + \langle \Sigma_{\rm CM}\rangle_{\TP}
       = 360\times 5 + 180 - 250 \sim 1730 \hbox{ MeV.}
\end{equation}
This estimate may look too crude because it ignores, for instance, the difference 
between the confinement of the 3-quark baryon and the pentaquarks, and also the other spin independent forces.  It has been shown, however, that a more realistic constituent quark model, such as the MIT bag model, gives similar values for the pentaquark masses.\cite{Carlson}
Thus the CM model requires an extra attraction of about 200 MeV  for $\TP$ ($1/2^-$).

The effect of FD on $\TP$ has been studied by Stancu and Riska\cite{SR} and also by Jennings and Maltman\cite{JM} and Bijker et al.\cite{Bijker}.
The latter two works also compare FD with CM.
It is found there that the FD interaction is slightly less repulsive. 

The instanton plays a crucial role also in $\TP$. 
The III (2-body) is strongly attractive, while the III (3-body) is slightly repulsive.
We find the $\TP$ mass employing the III to be the surprisingly small value
\begin{equation}
	   M_{\TP} = 5 m_q + \Delta m + \langle \Sigma_{\rm III}\rangle_{\TP}
       = 360\times 5 + 180 - 510 + 50 \sim 1520 \hbox{ MeV,}		
\end{equation}
where the large attraction ($-510$ MeV) comes from the 2-body III, and the last term,
the 50 MeV repulsion, is the contribution of the 3-body III.
Although this result is very close to the observed $\TP$ mass, this estimate may not be fully realistic.  
This is because the strong attraction is estimated perturbatively, assuming that the quark wave functions of the 3-quark baryon and the pentaquarks are the same.
A realistic estimate employing the MIT bag model\cite{STO} has been carried out and shows that the  
$\TP$ mass with the full III is about 1600 MeV.

In contrast to the situation described above, the positive parity baryon has a problem that it typically gives a mass
that is 100 MeV larger than that of the negative parity state.  In the CM case, we have
\begin{eqnarray}
	   M_{\TP} &=& 5 m_q + \Delta m + \langle \Sigma_{\rm CM}\rangle_{\TP}
	   +\Delta E(\Delta L=1)\nonumber\\
       && = 360\times 5 + 180 - 620 + 450\sim 1810 \hbox{ MeV.}	
\end{eqnarray}
The attraction due to the HF interaction is indeed strong, but it is cancelled by the kinetic energy of the $L=1$ orbital motion. The other models yield the same conclusion 
for $1/2^+$ as long as quark configurations based on the independent particle model, or the quark shell model, are considered.

The above exercises show that (1) the constituent quark model for the $1/2^-$ pentaquark tends to yield a larger mass than the observed $\TP$, 
though (2) the instanton induced interaction is a promising source of strong attraction for the pentaquark and 
(3) the $1/2^+$ state may not realize a lower energy than the $1/2^-$ state in the independent-particle quark models.

There are many model calculations that claim a positive parity state 
of lower energy than $1/2^-$.
It is, however, fair to say that the majority of such calculations contain an arbitrary parameter that is fitted to the observed $\TP$ mass or estimates the kinetic energy associated with 
$L=1$ unreasonably small.
One such example is the estimate in Ref.\cite{KL,JM}, which employ
$\Delta E (\Delta L=1) = 210 - 250$ MeV, based on the splitting of the $N^*$(1440) from the $N$(940).  But it is clearly unjustified to assume that the $N^*$(1440) is a 2 $\hbar \omega$ excited state and $\Delta E (\Delta L=1) =\hbar\omega$.

One possibility is to deviate from the independent particle picture.
Recently, a dynamical (but semi-classical) five-quark calculation\cite{Enyo} was carried out with a model Hamiltonian that contains a linear confinement potential plus the CM interaction. That calculation employs the antisymmetrized molecular dynamics (AMD) method, and the result shows that the $J^{\pi}= (1/2^+, 1/2^-) I=0$ and $J^{\pi}= (3/2^-) I=1$  states are almost degenerate as the lowest energy states.  The authors claim that the diquark-type correlation causes the $1/2^+$ state to have a smaller mass.  This is the subject of the next section.

\subsection{Diquarks}

In order to decrease the masses of the pentaquark states to the observed values in the quark model, 
Jaffe and Wilczek (JW)\cite{JW} proposed the diquark model.  A similar 
model was also proposed by Karliner and Lipkin (KL)\cite{KL}.

The diquark is a strongly correlated quark pair.  In QCD, both the gluon exchange interaction and the instanton induced interaction favor the spin-singlet and color-antisymmetric diquark combination,
\begin{equation}
{}[ud]_0 \quad	I = 0\quad C=\bar 3 \quad J^{\pi}=0^+.
\end{equation}
The instanton model calculation predicts a $[ud]_0$ diquark mass as small as 
420 MeV.

In the JW model, the structure of the $\TP$  is considered in terms of two diquarks and a strange antiquark system.
It is important that the color-singlet bound state of these three objects requires an orbital excitation, either to $L=1$ or to a state with a nodal radial wave function.
This can be understood from symmetry considerations: The two $[ud]_0$ should 
be antisymmetric in color so that a total color singlet state is obtained, while
they should be symmetric overall because they are ``bosons.'' 
Thus the relative motion between the diquarks should have $L=1$.
The mass of the $\TP$ is estimated as
\begin{equation}
M \sim 2M_D + m_s +\Delta E(\Delta L) \sim 840+560+ 450 = 1850 \hbox{ MeV.} 
\end{equation}

In the KL model,  the combination of a diquark and a triquark
$ud\bar s$ is considered, and the $\TP$ mass is estimated.  
It is also assumed that the relative
angular momentum is $L=1$, but the estimated value of $\Delta E(\Delta L) $ is much
smaller, $\sim 209$ MeV.

How realistic are the diquarks? An attempt was made in lattice QCD to calculate the diquark mass.  As the diquark is not color singlet, it is necessary to fix the gauge to calculate the mass.
Wetzorke and Karsch\cite{WK} used the Landau gauge and derived the spectrum of the diquarks. A recent calculation employing the maximal entropy method for the diquark spectral function yields the conclusion that the mass of the $[ud]_0$ diquark is about 600 MeV.
There is also an approach using the Bethe-Salpeter equation within rainbow-ladder QCD\cite{Maris} that predicts a diquark mass $\sim 800$ MeV.

We point out that the diquark picture for the pentaquark system requires reexamination of the nucleon and other ground state baryons with the same footing.
An interesting quantity is the magnetic moments of the ground state baryons. Because the diquark enhancement may break the $SU(6)$ symmetry, the relations among the magnetic moments of the octet baryons may not hold.
For instance, the ratio of the magnetic moments of the proton and neutron is given 
in the diquark limit by
\begin{equation}
\mu_p /\mu_n = e_u /e_d = -2
\end{equation}
if the quarks do not have anomalous moments.

\bigskip
In summary, the situation is not at all clear in the ``constituent'' quark model.
In fact, the quark model has been tested in the case of the nonexotic mesons and baryons, but it may be necessary to modify it to accommodate the pentaquarks and other exotic states.
The suggested strong correlation realized in the diquarks, if it is real, may
make it necessary to reconsider the basic dynamics of the constituent quark model.
The width is another quantity that is to be explained in the quark model. 
The ``fall-apart'' theory is conventionally employed and predicts a width of the pentaquarks
that is too large,
but the validity of such a theory has not yet been confirmed.

\section{QCD prediction}

In the situation that the quark model cannot provide definite predictions for the spin, parity and structure of the pentaquark state, we desperately need a QCD calculation.
Several attempts have been made to apply QCD directly to this problem. 

\subsection{QCD sum rules}
The QCD sum rule\cite{SVZ} has been applied to the $\TP$ spectrum by Zhu\cite{Zhu}, 
Matheus et al.,\cite{Matheus} and Sugiyama et al.\ (SDO)\cite{SDO}.
The first two approaches consider only the chiral even terms of the
sum rule and do not consider the parity projection.

SDO made the first attempt to determine the parity and mass of $\TP$ from the QCD sum rule, assuming that its spin is $\half$.
The QCD sum rule is derived by computing the relevant correlation function
in two ways, with the operator product expansion (OPE) in the deeply Euclidean region, and with a phenomenological parameterization of the spectral function.
This sum rule relates hadron properties directly to the QCD vacuum condensates, such as $\qq$ and $\GG$, as well as to the other fundamental constants, such as $m_s$.  

In Table II, possible local operators for the pentaquark systems are listed.
We employ the following interpolating field operator for $\TP$:
\begin{eqnarray}
  J(x)&=&\epsilon_{abc}
   \bar S_a S_b^5 C\bar{s}_g^T(x) .
   \label{eq:IF}
\end{eqnarray}
Here, $a,b,c,\cdots$ are color indices and $C=i\gamma^2\gamma^0$.
Also, $S_c^5(x) = \epsilon_{abc} u_a^T(x)C\gamma_5 d_b(x)$ and
$S_c(x) = \epsilon_{abc} u_a^T(x)Cd_b(x)$ are the scalar ($0^+$) and the pseudoscalar ($0^-$) $ud$ diquark operators. They both belong to color $\bar 3$ and $I=0$.
The operator $J(x)$ produces a baryon with $J=\half$, $I=0$ and strangeness $+1$.
One of the advantages of $J(x)$ is that its coupling to the main continuum state, $NK$, is believed to be small, because it cannot be decomposed 
into a product of $N(3q)$ and $K(q\bar q)$ operators in the nonrelativistic limit.

It is important that the correlation function of $J(x)$ contains both positive and negative parity components.
Therefore it is necessary for
the parity projection to determine the parity of the pentaquark state.
Such a technique was developed by Jido et al.\cite{JKO} and
it has been applied to the baryon
excited states.
Here, we follow the same procedure for the pentaquark correlation function.

We consider the retarded Green's function and choose the rest frame, $\vec q=0$, 
\begin{equation}
 \Pi (q_0) =  \int d^4 x\,  e^{iq\cdot x} i \bra 0| \theta(x^0) \,J(x) \bar{J}(0) |0\ket|_{\vec q=0} \, .
 \label{eq:Pi}
\end{equation}
Then the imaginary part of $\Pi$ contains two functions, which are the sum and difference of
the positive and negative parity spectral functions, $\rho^{\pm}$,
\begin{eqnarray}
\label{eq:AB}
 \frac{1}{\pi}{\rm Im} \Pi(q_0) &= & A(q_0) \gamma^0 + B(q_0) \\
 A(q_0) &=&  \frac{1}{2} \left(\rho^+(q_0) + \rho^-(q_0)  \right)\nonumber\\
 B(q_0) &=&  \frac{1}{2} \left(\rho^+(q_0) - \rho^-(q_0)  \right)  ,\nonumber
\end{eqnarray}
or equivalently,
\begin{eqnarray}
 \rho^{\pm}(q_0)  &=& A(q_0) \pm B(q_0) .
 \label{eq:rhopm}
\end{eqnarray}
It happens that the part $B$ in the OPE contains only the chiral-odd terms, 
such as $\qq$ and $\qGq$, 
and thus the chiral symmetry breaking is responsible for the parity splitting.

The sum rule is obtained by comparing the OPE of the correlation function and 
explicit forms of the spectral functions using analytic continuation.
The spectral function is commonly parametrized as a pole plus the continuum contribution:
\begin{eqnarray}
    \rho^{\pm}_{\rm Phen}(q_0) &=&  {|\lambda_{\pm}|^2 } \delta(q_0 - m_{\pm}) 
                 +\theta(q_0- \sth)  \rho^{\pm}_{\rm CONT}(q_0) .
\label{eq:rho}
\end{eqnarray}
Here, the continuum part is assumed to be
identical to the corresponding OPE function above the threshold $\sth$.
In order to both enhance the pole part and also suppress the higher-dimensional terms of the OPE,
we introduce a weight function 
of the ``Borel'' type.
Then the sum rule is obtained as
\begin{equation}
\int dq_0 W(q_0) \rho^{\pm}_{\rm Phen} (q_0)= \int dq_0 W(q_0) \rho^{\pm}_{\rm OPE} (q_0) .
\end{equation}
with
$W(q_0) = \exp \left( - \frac{q_0^2}{\MB^2} \right)$,
where $\MB$ is the relevant (Borel) mass scale for the baryon.  
The values of the QCD parameters, $m_s$, $\sbs$, $\sGs$ and $\GG$, are chosen so that the
strange baryon masses are reproduced in the sum rule.
The values we choose are as follows:
$m_s = 0.12 $ GeV,  $\sbs = 0.8 \times (-0.23~ {\rm GeV})^3$,
$m_0^2 \equiv \sGs / \sbs = 0.8~{\rm GeV}^2$ and
$\GG = (0.33~{\rm GeV})^4$.

The residue of the pole, $|\lambda_{\pm}|^2$,  in Eq.~(\ref{eq:rho}) represents
the strength with which the IF couples to the physical state, and it
should be positive only when the pole is real.
We use this condition to determine the parity of the pentaquark state.  
In Fig.~3, we plot the OPE side (as a function of $\MB$) corresponding to
$|\lambda_{\pm}|^2 \exp (-m^2/\MB^2)$.
We find that the dimension-5 condensate, $\sGs$, gives a large negative contribution to 
$|\lambda_{+}|^2$, which ends up with a nearly zero or even slightly negative value.  
This suggests that the pole in the positive-parity spectral function is spurious.
In contrast, the large $\sGs$ contribution makes $|\lambda_{-}|^2$ positive, and 
we thus conclude that
the obtained negative-parity state is a real state.

\begin{figure}[hbtp]
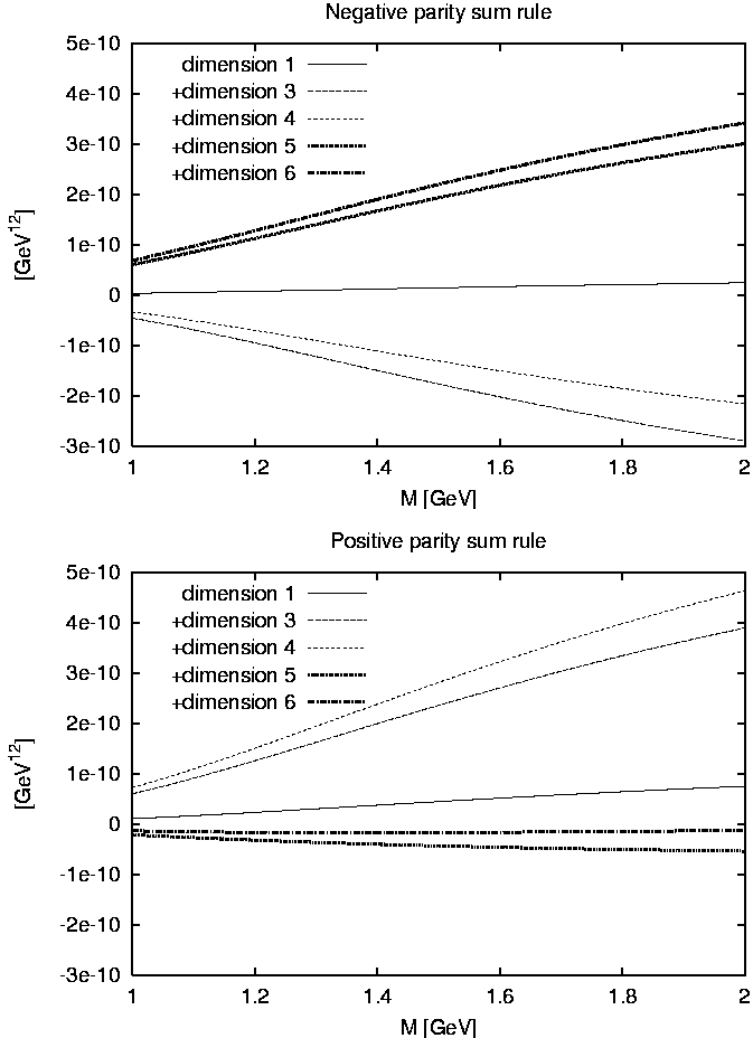

\begin{center}
\includegraphics*[width=10cm]{tneg.epsf}
\includegraphics*[width=10cm]{tpos.epsf} 
\caption{
Contributions from the terms of each dimension added up successively
for the negative-parity and positive-parity sum rules with $\sth=1.8$ GeV.%
}
\end{center}
\label{fig:R}
\end{figure}

The mass of the negative parity $\TP$ is estimated using the sum rule, and we find
that the $\MB$ dependence is rather weak. Therefore the reliability of the sum rule 
seems good.
It is, however, sensitive to the choice of the threshold, and the result is ambiguous to some extent.  We have only confirmed that the result is consistent with the observed value.

Summarizing this part, the QCD sum rule shows that (1) the parity splitting is caused by chiral symmetry breaking due to $m_s$,  $\qq$ and $\qGq$ condensates.
(2) The parity of $\TP$ is determined by the positivity of the correlation function and
the negative parity state is predicted.
(3) The mass of the $1/2^- \TP$ baryon is obtained as
$1.3 - 1.7$ GeV.

An important issue regarding the QCD sum rule is the choice of the interpolating field operator.  If the IF is inappropriate, the sum rule does not
contain the pentaquark contribution, but the $KN$ continuum state may dominate.
A recent study\cite{Kondo} indicates that the continuum effect on the sum rule might  change the conclusion concerning the parity assignment.  Further studies on the choice of the IF and the sum rule approach are necessary.

\subsection{Lattice QCD}

Several lattice QCD calculations suggest that the positive parity $\TP$
is very massive.
Csikor et al.,\cite{CFKK} Sasaki\cite{Sasaki} and Lee et al.\cite{Kentucky} calculated
the $\TP$ correlation function in quenched lattice QCD.
They employed different local interpolating field (IF) operators that 
do not contain derivatives. They are independent, but  
from the Fierz transform, it is seen that they are not orthogonal.

Lee et al.\ have used the IF operator given by the product of $N$ and $K$,
i.e., $\TP_{NK}$ in Table II.  They concluded that the correlator is consistent with
the $NK$ background and exhibit a resonance state in neither $1/2^-$ nor $1/2^+$.
This result can be understood naturally, because their IF has a large overlap with the
$NK$ continuum state.

In contrast, in both Refs. \cite{CFKK} and \cite{Sasaki}, it is claimed that there exists a resonance state of $1/2^-$ at a mass consistent with the observed $\TP$.
Sasaki employed the same IF as that used in the QCD sum rule in the previous section,
and he found that the effective mass extracted from the  
time dependences of the correlation function possesses two plateaus. 
One corresponds to the $NK$ threshold and the other indicates 
a narrow resonance above the $NK$ threshold. 
It is important to confirm whether such a double
plateau structure is not an artifact of lattice QCD.

The Hungarian group\cite{CFKK} used another IF and found that $\TP$ has
$J^{\pi}=1/2^-$ and that its mass is consistent with the $NK$ threshold. It is not
clear whether they can claim a resonance above the threshold or not.

It is fair to say that the majority of the calculations agree that the ground state is
characterized by
$J =1/2$, $I =0$, and negative parity, while the positive parity state has a mass that is greater at least by a few hundred MeV.
These calculations also show that the results obtained to this time do not depend 
strongly on the choice of the interpolating field.

\section{Conclusion}

The conclusions of this article are itemized in the following.
\begin{itemize}
\item QCD predicts the existence of a $\TP$ ($J^{\pi}=1/2^-$, $I =0$), 
although it may not be 
easy to distinguish $\TP$ from the $KN$ threshold in the sum rule or in lattice QCD.
\item Many models suggest positive parity baryons.
The soliton models provide a parametrization of the masses of the ``pentaquark'' states
that is consistent with the observed masses.
The quark models for the positive parity states seem to often use a kinetic excitation energy
that is too small,
$\Delta M = M(L=1) - M(L=0) \sim 200$ MeV, while a reasonable estimate is about 
$400-450$ MeV.
Unless an exotic idea, such as a strong diquark correlation, is introduced, 
it seems unlikely that the $1/2^-$ state will have a higher energy than the $1/2^+$ state in the quark model.
\item 
The quark model predicts a mass of $\TP$ that is larger than the observed value by about 100 MeV - 200 MeV.
The discrepancy is larger for $1/2^+$.  
The idea of the strong diquark correlation may require confirmation from the QCD viewpoint.
\item The real challenge is to account for the small width.  It was even pointed out that consistency with the existing $KN$ phase shift analyses necessitates a tiny width, 
as small as 0.1 MeV, in particular for the negative parity $\TP$.\cite{Arndt} 
A mechanism yielding strong suppression of the decay is desperately needed.
\end{itemize}

New experiments with better statistics and with different production mechanisms have already appeared and will continue to appear. 
Most theoretical models predict 
not one but in fact many pentaquark states, and therefore observations of other states are extremely important.  Among them, the spin $J=3/2$ partner and the other members of the antidecuplet, as well as the octet, which may be mixed with $\tenbar$, are particularly anticipated.

Recently, new theoretical papers on the pentaquark appear daily.  We need further study of the width and the production mechanisms as well as realistic (not toy-model) calculations based on QCD.  Lattice QCD with improved statistics and other interpolating field operators is also necessary to confirm the nature of pentaquarks.

It would also be fascinating to determine whether pentaquark spectroscopy is
related to other newly-observed narrow hadron states, such as $D_s^*$ and $X(3872)$.
It would be interesting to study whether the heavy quark plays a key role in these new resonances.

\section*{Acknowledgments}
This article is based on the talk given at the YITP Workshop on ``Multi-quark Hadrons; Four, Five and More?'', held on Feb.\ 17-19, 2004.
The hospitality of the Yukawa Institute for Theoretical Physics, Kyoto University is deeply appreciated. I also thank Dr.\ Teiji Kunihiro for encouraging me to write this article. 
I would like to thank Drs. S.~Takeuchi, T.~Doi, T. Shinozaki and J.~Sugiyama for 
the collaborations from which a part of the content of this article has been produced.
I also acknowledge Profs. T.~Nakano, K. Hicks, A.~Hosaka, R.L.~Jaffe, 
O.~Morimatsu, Y.~Kondo for fruitful discussions.

%

\end{document}